\documentclass[10pt,prl,letterpaper,bibnotes,balancelastpage,notitlepage,twocolumn,floatfix,superscriptaddress,nofootinbib,preprintnumbers,amsmath,amssymb,aps]{revtex4-2}

\usepackage{graphicx}
\usepackage{dcolumn}
\usepackage{bm}
\usepackage{microtype}
\usepackage{xcolor}
\definecolor{navyblue}{rgb}{0.0, 0.0, 0.5}
\usepackage{hyperref}
\hypersetup{
    colorlinks=true,
    linktoc=all,
    allcolors=navyblue
}

\renewcommand{\section}[1]{\phantomsection\addcontentsline{toc}{section}{#1}\textbf{\textit{#1}.}---\unskip\ignorespaces}

\usepackage[capitalize]{cleveref}
\crefname{section}{Sec.}{Secs.}
\Crefname{section}{Sec.}{Secs.}
\crefname{appendix}{App.}{Apps.}
\Crefname{appendix}{App.}{Apps.}
\Crefname{figure}{Fig.}{Figs.}
\crefname{figure}{Fig.}{Figs.}
\creflabelformat{equation}{#2#1#3}

\usepackage{color,environ}
\usepackage{fontawesome5}
\definecolor{orcidlogocol}{rgb}{0.65, 0.807, 0.223}
\newcommand{\orcid}[1]{$\,$\href{https://orcid.org/#1}{\textcolor{orcidlogocol}{\footnotesize\faOrcid}}}

\graphicspath{{./}{./}}

\newcommand*\R{\mathbb R}

\newcommand*\te[1]{\text{#1}}

\newcommand*\p[1]{\left(#1\right)}
\newcommand*\ps[1]{\left[#1\right]}
\newcommand*\pc[1]{\left\{#1\right\}}
\newcommand*\f[2]{\frac{#1}{#2}}
\newcommand*\mat[2]{\left(\begin{array}{#1}#2\end{array}\right)}

\newcommand*\I{\te{i}}

\newcommand{\TQCD}{T_{\rm QCD}}
\newcommand{\gagg}{g_{a \gamma \gamma}}
\newcommand{\md}{m_S}
\newcommand{\fd}{f_S}
\newcommand{\ma}{m_a}
\newcommand{\maz}{m_{a,0}}
\newcommand{\fa}{f_a}
\newcommand{\phid}{\phi_S}
\newcommand{\phia}{\phi_a}
\newcommand{\phih}{\phi_H}
\newcommand{\phil}{\phi_L}
\newcommand{\Tcross}{T_\times}
\newcommand{\tcross}{t_\times}
\newcommand{\Thzd}{\Theta_{0,S}}
\newcommand{\Tosc}{T_{\rm osc}}
\newcommand{\mpl}{m_{\mathrm{pl}}}

\NewEnviron{eq}{%
\begin{align}\begin{split}
  \BODY
\end{split}\end{align}
}

 \definecolor{mulberry}{rgb}{0.5,0,0.5}
 \definecolor{elderberry}{rgb}{0.3,0,0.7}

\usepackage[normalem]{ulem}
\usepackage{mathtools}  

\makeatletter
\newcommand{\amarki}{\faCoffee}
\newcommand{\amarkii}{\faPaperPlane[regular]}
\def\@fnsymbol#1{{\ifcase#1\or \amarki\or \amarkii\or \amarkiii\or \amarkiv \else\@ctrerr\fi}}
\makeatother

\begin{document}
\raggedbottom


\title{Heavy QCD axion dark matter from avoided level crossing}

\author{David Cyncynates\orcid{0000-0002-2660-8407}}
\email{davidcyn@uw.edu}
\affiliation{Department of Physics, University of Washington, Seattle, WA 98195, U.S.A.}

\author{Jedidiah O. Thompson\orcid{0000-0002-7342-0554}}
\email{jedidiah@stanford.edu}
\affiliation{Stanford Institute for Theoretical Physics, Stanford University, Stanford, CA 94305, U.S.A.}

\date{\today}

\begin{abstract}
The QCD axion offers a natural resolution to the strong CP problem and provides a compelling dark matter candidate. If the QCD axion constitutes all the dark matter, the simplest models pick out a narrow range of masses around $100\,\mu{\rm eV}$. We point out a natural production mechanism for QCD
axion dark matter at masses up to existing astrophysical bounds ($\mathcal{O}(20 \, \mathrm{meV})$ for the most minimal models and $\mathcal{O}(1 \, \mathrm{eV})$ for nucleophobic models).
If the QCD axion mixes with a sterile axion, the relative temperature dependence of their potentials can lead to an avoided level crossing of their mass eigenstates. This leads to a near-total transfer of energy density from the sterile axion to the QCD axion, resulting in a late-time QCD axion abundance sufficient to make up all of present-day dark matter.  Our result provides additional theoretical motivation for several direct detection experiments that will probe this part of parameter space in the near future.
\end{abstract}

\maketitle

\section{Introduction} \label{sec:Intro}
The QCD axion is one of the best-motivated candidates for physics Beyond the Standard Model (BSM). Although it was originally proposed as a solution to the strong-CP problem~\cite{Peccei:1977hh,Weinberg:1977ma,Wilczek:1977pj}, it was quickly realized that such a new field could also have cosmological consequences~\cite{Preskill:1982cy,Abbott:1982af,Dine:1982ah,Duffy:2009ig}. In particular, it is an excellent candidate for dark matter (DM), for which we have overwhelming evidence from a number of sources \cite{Bertone:2004pz,Garrett:2010hd}.

At low energies and in the absence of further model-building, the properties of the QCD axion are determined (apart from some $\mathcal{O}(1)$ model dependencies) entirely by one parameter: its zero-temperature mass $\maz$. The axion has several couplings to the Standard Model (SM) whose strengths are typically set by its decay constant $\fa \sim m_\pi f_\pi / \maz$ where $m_\pi f_\pi \sim 200{\rm \,MeV}$. One particularly promising coupling to target is the axion-photon coupling $\mathcal{L} \supset - \frac{\gagg}{4} \phia F_{\mu \nu} \tilde{F}^{\mu \nu}$, where $\phia$ is the axion field and $F_{\mu \nu}$ is the SM photon field strength. In minimal models, the constant $\gagg$ is given by $\gagg = C_{a \gamma \gamma} \alpha_{\rm QED} / 2 \pi \fa$ where $C_{a \gamma \gamma}$ is an $\mathcal{O}(1)$ model-dependent constant and $\alpha_{\rm QED}$ is the fine-structure constant.

An axion making up the entirety of DM is ruled out for masses $\maz \gtrsim 20 \, \mathrm{meV}$, since the axion coupling to SM nucleons would lead to anomalous energy loss in neutron stars \cite{Buschmann:2021juv} and SN1987A \cite{Carenza:2019pxu}. In models where the QCD axion couples only weakly to nucleons (so-called nucleophobic models \cite{DiLuzio:2017ogq,Badziak:2023fsc}), the above axion-photon coupling still places a bound of $\maz \lesssim 1 \, \mathrm{eV}$~\cite{Ayala:2014pea,Dolan:2022kul}.
At lower masses, however, there are few phenomenological constraints on axion DM, and in fact there are many existing and planned experiments that are probing regions here \cite{IAXO:2019mpb,Baryakhtar:2018doz,Schutte-Engel:2021bqm,BREAD:2021tpx,BRASS,Aja:2022csb,McAllister:2017lkb,Beurthey:2020yuq,Lawson:2019brd,DeMiguel:2023nmz,Stern:2016bbw,DMRadio:2022pkf,Berlin:2020vrk}.
It is thus important to understand the possible production mechanisms for axion DM, as they guide the most well-motivated search targets. Since the axion arises as a pseudo-Nambu-Goldstone boson associated with a new Peccei-Quinn (PQ) symmetry, these mechanisms split into two categories depending on whether this symmetry is broken before or after inflation ends.

If PQ symmetry is broken after inflation, then the axion takes random initial values in each Hubble patch at the time of breaking. This stochastic initial field evolves into a complicated network of axion strings and domain walls that collapse and decay around the time of the QCD phase transition \cite{Davis:1986xc}. 
These dynamics can be simulated, and the simulations may be used to extract a sharp prediction for the mass of a post-inflationary axion: $\maz \sim 40\div 180\,\mu{\rm eV}$ \cite{Buschmann:2021sdq}. 
Although there is some modeling uncertainty \cite{Gorghetto:2020qws,Gorghetto:2018myk}, it seems unlikely that  post-inflationary production could yield an eV-scale mass.

If PQ symmetry is broken before the end of inflation on the other hand, then the axion initial field value is effectively homogeneous and, depending on the inflationary history, non-zero across the observed universe. This is the misalignment mechanism; the initial value is known as the axion misalignment angle $\Theta_0$, and it (along with the axion mass) determines the present-day axion energy density. Because the QCD axion potential is periodic, $\Theta_0$ is valued in the range $[-\pi , \pi )$. One minimal possibility is therefore that $\Theta_0 \sim \mathcal{O}(1)$, as would be the case if it were selected by UV dynamics insensitive to the low-energy QCD potential. For $|\Theta_0| \sim \pi/2$, the present-day DM abundance is produced for a QCD axion with mass $\maz \sim 10\,\mu{\rm eV}$. For axion masses smaller than this, the misalignment mechanism requires $|\Theta_0| \ll 1$, which can be explained either by dynamic \cite{Co:2018phi,Papageorgiou:2022prc} or anthropic \cite{Linde:1987bx,Tegmark:2005dy,Freivogel:2008qc} arguments. However for much heavier axion masses $\maz \gg {\rm meV}$, it becomes significantly more difficult to produce the proper DM abundance via misalignment. QCD axions with masses $\maz \sim 1 \, \mathrm{eV}$ for example would require an initial misalignment angle tuned extremely close to $\pi$: $\pi - |\Theta_0| \approx e^{-10^3}$ \cite{Marsh:2015xka,Arvanitaki:2019rax}. Such a possibility is not only aesthetically problematic, but also violates inflationary isocurvature constraints \cite{Arvanitaki:2019rax}.

On the other hand, this high-mass region of parameter space is experimentally interesting, with several new experiments either probing or set to probe QCD axion dark matter with masses $0.2{\,\rm meV} \lesssim \maz \lesssim 1{\,\rm eV}$ \cite{Baryakhtar:2018doz,BREAD:2021tpx,Schutte-Engel:2021bqm}. It is thus important to understand what type of model can produce heavy QCD axion DM and  how complicated such a model must be. 

The landscape of such models is far from vacant. There exist a handful of models that rely on various dynamics of a UV partner field (often taken to be the radial mode of the complex axion parent field) to generate energy density that subsequently converts into axions, either via parametric resonance \cite{Co:2017mop} or intricate non-thermal dynamics \cite{Harigaya:2019qnl,Co:2019jts}. In the post-inflationary scenario, there is also the possibility that the QCD axion has multiple nearly-degenerate vacua, which can lead to the enhanced late-time axion energy density necessary to be DM \cite{Kawasaki:2014sqa,Hiramatsu:2010yn,Hiramatsu:2012sc,Ringwald:2015dsf,Harigaya:2018ooc,Caputo:2019wsd}.
In this letter, we will describe a different type of model that is both simple and free of any tuning problems. The dynamics we discuss are also quite independent of any new couplings of the axion in the UV.

The key insight is as follows. It is possible (indeed even theoretically well-motivated \cite{Arvanitaki:2009fg}) that there are one or more additional axions in the theory. We will consider one such field $\phid$ with mass $\md$ and decay constant $\fd$, which we will call ``sterile'' as $\fd\gtrsim\fa$ in the parameter space of interest. For such a field, $\md$ and $\fd$ can be effectively independent, and if it is sourced by misalignment, then the present-day energy density is generically proportional to $\fd^2$. However it turns out that this energy density can be easily transferred to the QCD axion via the temperature-dependent nature of the QCD axion potential. The QCD axion mass is extremely small at high temperatures but increases to its zero-temperature mass $\maz$ as the universe cools below the QCD scale. It is thus possible that the QCD and sterile axion masses cross each other, and if there is any interaction between these two fields then the mass eigenstates can instead undergo an avoided crossing, leading to an adiabatic transfer of energy from $\phid$ to $\phia$.\footnote{A related phenomenon is pointed out in Ref.~\cite{Ho:2018qur}, wherein the QCD axion transfers its energy to the sterile axion.} The sterile axion thus acts effectively as a battery that stores enough energy density for the QCD axion to become the dark matter at late times.

\section{Dynamics} \label{sec:Dynamics}
We consider the following model of the QCD axion $\phia$ interacting with a sterile axion $\phid$ (which we UV complete in the subsequent section):
\begin{equation} \label{eq:lagrangian}
\begin{aligned}
    {\mathcal L} &\supset \f12(\partial\phia)^2 + \f12(\partial\phid)^2 \\&- \ma^2(T)\fa^2\ps{1 - \cos\p{\f{\phia}{\fa} + \f{\phid}{\fd}}} \\&- \md^2 \fd^2\ps{1 - \cos\p{\f{\phid}{\fd}}}\,,
\end{aligned}
\end{equation}
where we approximate the temperature dependence by the simplified expression:
\begin{align}
\ma^2(T) = \maz^2 \max\pc{1,\p{\f{T}{T_{\rm QCD}}}^{-n}}\,,
\end{align}
with $T_{\rm QCD}\approx 100 {\rm MeV}$ and $n \approx 6.68$ in the dilute instanton gas approximation \cite{Wantz:2009it}.
The interesting dynamics that we will study will occur when $\md \ll \maz$ and $\fd \gg \fa$, and from here forward we will work in this region of parameter space.

At leading order, we may approximate the two-axion potential by its quadratic terms
\begin{align}
    V\approx\f12\mat{cc}{\phia&\phid}\left(
\begin{array}{cc}
 \ma^2 & \frac{\fa }{\fd}\ma^2 \\
 \frac{\fa }{\fd}\ma^2 & \md^2+\frac{\fa^2 }{\fd^2}\ma^2 \\
\end{array}
\right)\mat{c}{\phia\\\phid}\,.
\end{align}
The fields $\phia, \phid$ are thus not propagation eigenstates. Instead, we must rotate to a basis in which this mass matrix is diagonalized. 
As the temperature of the universe drops and $\ma(T)$ evolves, the propagation basis rotates, which leads to adiabatic energy density transfer between the two fields.

If $\fd \gg \fa$, the mass matrix is nearly diagonal. At early times, $\ma(T) \ll \md$ and we can find the heavy and light mass eigenstates to be $\phih \approx \phid$ and $\phil \approx \phia$ respectively. At late times the temperature has dropped, leading to $\ma(T) \gg \md$. Now the heavy and light mass eigenstates are given by $\phih \approx \phia$ and $\phil \approx \phid$ respectively. At some intermediate temperature, when $\ma(T) \approx \md$, the mass matrix is nearly the identity and the two mass eigenvalues are nearly degenerate. However the off-diagonal elements split this degeneracy and lead to an avoided crossing of the eigenvalues. Provided the transition through this avoided crossing is adiabatic (meaning slow compared to the oscillatory timescale of the two axion fields), all energy density contained in the heavy propagation eigenstate will remain in the heavy eigenstate. In other words, energy density will be smoothly transferred from $\phid$ to $\phia$.

We will now check the necessary condition for this transition to be adiabatic, and then compute the present-day energy density in the QCD axion after these early-time dynamics. The avoided crossing occurs at a time when the mass matrix is approximately the identity, so we define the crossing temperature $\Tcross$ and time $\tcross$ to occur when $\ma^2 = \md^2 + \frac{\fa^2}{\fd^2} \ma^2 \approx \md^2$. The timescale over which the crossing happens is set by when the off-diagonal terms are important. A parametric estimate is that it begins when $\ma^2 - \md^2 \approx \frac{\fa}{\fd} \ma$ and ends when $\ma^2 - \md^2 \approx - \frac{\fa}{\fd} \ma$. From this we can calculate that the crossing lasts for a parametric duration $\Delta \tcross$ given by:
\begin{equation}
    \Delta \tcross\approx \f{3}{n}\f{\fa}{\fd} \sqrt{\f{20}{\pi^3 g_\star(\Tcross)}}\f{m_{\rm pl}}{T_{\rm QCD}^2}\p{\f{\md}{\maz}}^{4/n}\,,
\end{equation}
where $m_{\rm pl} = G_N^{-1/2}$ is the Planck mass.
In order for the transition to be adiabatic, we require:
\begin{equation} \label{eq:adiabaticityCondition}
    \Delta \tcross \gg \md^{-1} \approx \ma(\Tcross)^{-1}\,,
\end{equation}
which can easily be satisfied so long as $\fd$ is not too large.

The equations of motion satisfied by the homogeneous axion field are identical to the equations satisfied by the spatial axion perturbations, up to a common diagonal gradient term \cite{Cyncynates:2021xzw}. An avoided crossing will therefore also occur for each spatial fourier mode, causing the QCD axion to inherit the perturbations of the sterile axion. So long as the sterile axion is initialized with negligible isocurvature, the QCD axion will have purely adiabatic perturbations.

Now provided the transition is adiabatic, nearly all of the energy density in $\phid$ before the crossing will be transferred to $\phia$ after the crossing. We can thus estimate the late-time QCD axion abundance as follows. The initial energy density in the sterile axion field before it begins oscillating is given by $\rho_S(H \gg \md) \approx (1/2) \md^2 \fd^2 \Thzd^2$, where $\Thzd$ is the initial misalignment angle of $\phid$. At a time $H \sim \md$, the sterile axion field starts oscillating and this energy density begins redshifting as $a^{-3}$ (where $a(t)$ is the scale factor). At level crossing, this energy density is transferred to the QCD axion, but by construction this must happen at a time when the QCD axion mass is still below its zero-temperature mass. As $\ma(T)$ increases, the energy density stored in the QCD axion field also increases, going as $a^{(n-6)/2}$. Finally, when the QCD axion reaches its zero-temperature mass (i.e.\ when the universe temperature is $T \sim \TQCD$), this energy density again starts redshifting as $a^{-3}$ and does so until the present day. Putting all of this together, we obtain an estimate for the present-day QCD axion energy abundance:
\begin{align}\label{eq:relicAbundance}
    \Omega_a \approx \frac{4 \pi}{3} \frac{\maz \md \fd^2 \Thzd^2}{\mpl^2 H_0^2}\f{a^3(T_{\rm osc})}{a^3(T_{0})}\,,
\end{align}
where we have defined the oscillation temperature by $3 H(\Tosc) = \md$. 
\begin{figure}
    \centering
    \includegraphics[width = \columnwidth]{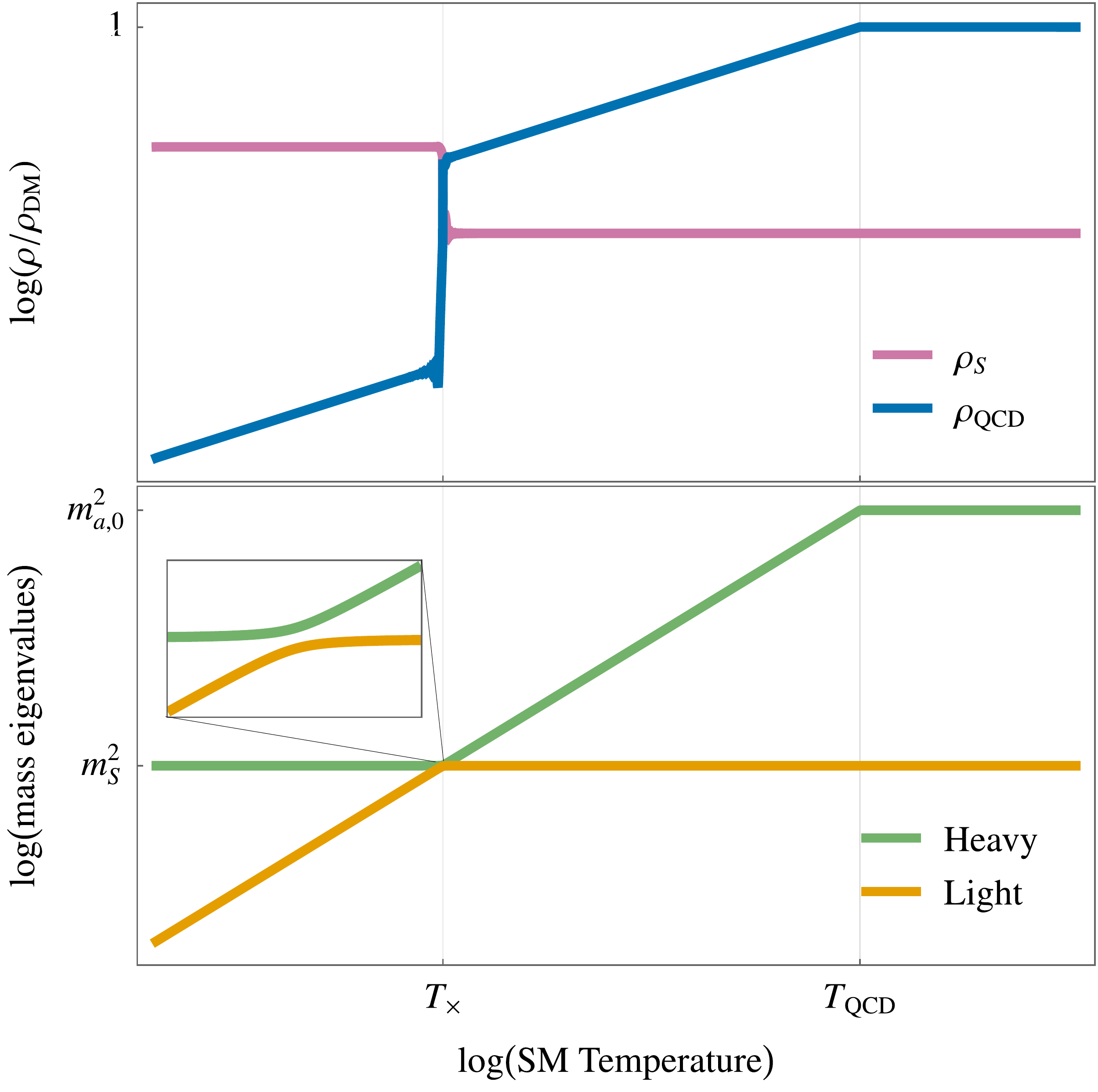}
    \caption{(Upper panel) In pink and blue, we plot the energy density of the mass eigenstate that is most strong coupled to the sterile and QCD instantons respectively. Note that at crossing, the two states are roughly equally coupled to the QCD instanton, but away from crossing this distinction is robust. (Lower panel) In green and orange, we plot the heavy and light mass eigenvalues respectively. The inset shows the avoided crossing in detail. In both plots, temperature decreases (and time increases) to the right.}
    \label{fig:Evolution}
\end{figure}
\begin{figure}
    \centering
    \includegraphics[width = \columnwidth]{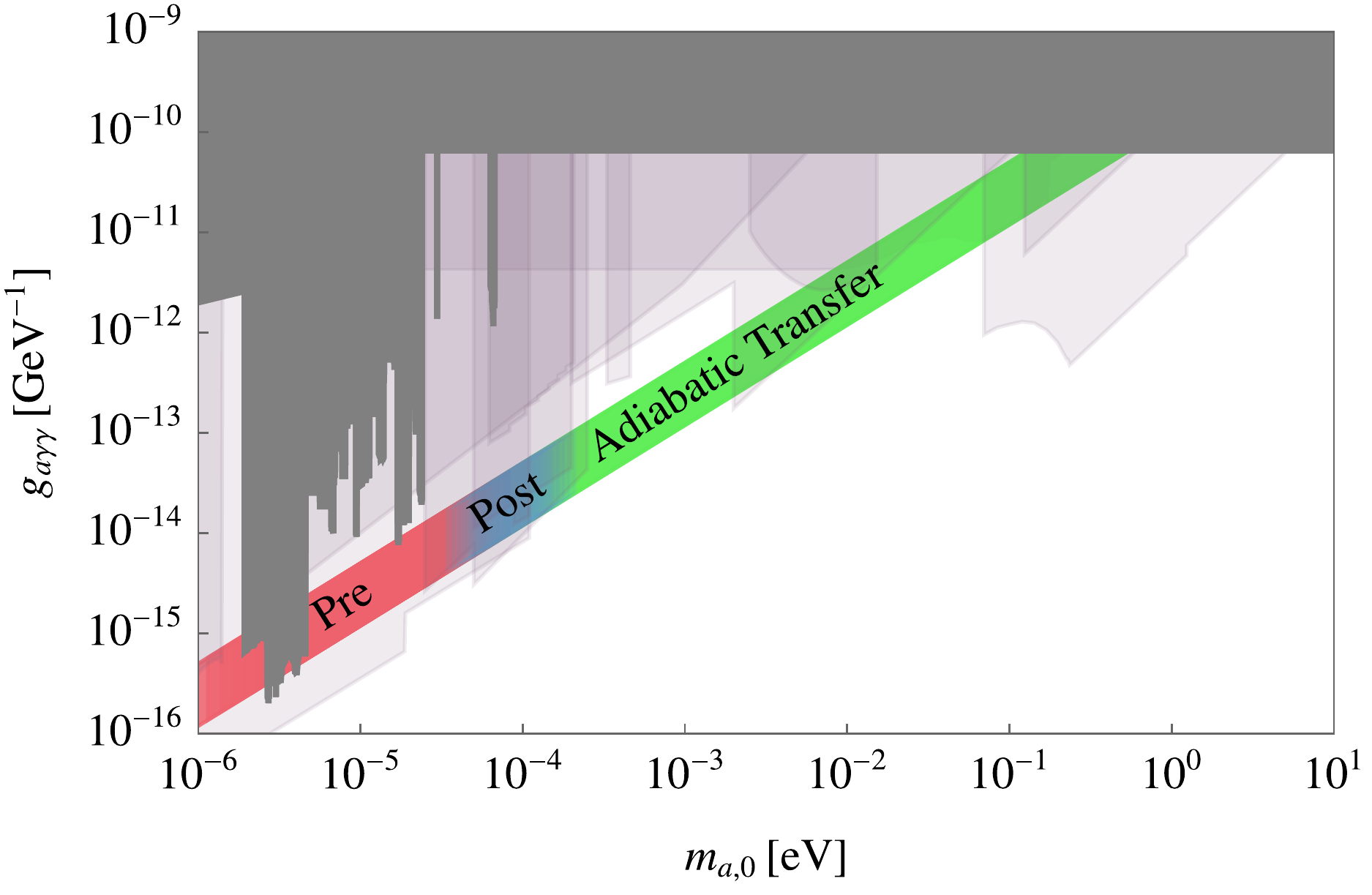}
    \caption{The axion-photon coupling $\gagg$ versus the axion mass $m$. The QCD axion line is highlighted in three colors, corresponding roughly to the range of massess accessible to three different production mechanisms. In red is pre-inflationary production  assuming an initial misalignment angle in the range $[0.1\pi,0.9\pi]$ \cite{GrillidiCortona:2015jxo,Marsh:2015xka}, in blue is post-inflationary production, where the mass range is taken from Ref.~\cite{Buschmann:2021sdq}, and in green are the higher masses accessible through adiabatic transfer (this work). The dark gray regions are excluded either by dark matter haloscopes \cite{ADMX:2018gho,ADMX:2018ogs,ADMX:2019uok,Crisosto:2019fcj,ADMX:2021nhd,ADMX:2021mio,Lee:2020cfj,Jeong:2020cwz,CAPP:2020utb,Lee:2022mnc,Kim:2022hmg,Yi:2022fmn,Adair:2022rtw,DePanfilis:1987dk,Hagmann:1990tj,HAYSTAC:2018rwy,HAYSTAC:2020kwv,HAYSTAC:2023cam,Alesini:2019ajt,Alesini:2020vny,Alesini:2022lnp,Quiskamp:2022pks} or by astrophysical probes \cite{Ayala:2014pea,Dolan:2022kul,Noordhuis:2022ljw}. The light-gray transparent regions are prospective sensitivity curves for upcoming experiments \cite{IAXO:2019mpb,Baryakhtar:2018doz,Schutte-Engel:2021bqm,BREAD:2021tpx,BRASS,Aja:2022csb,McAllister:2017lkb,Beurthey:2020yuq,Lawson:2019brd,DeMiguel:2023nmz,Stern:2016bbw,DMRadio:2022pkf,Berlin:2020vrk}. The data used to make this plot is compiled in Ref.~\cite{AxionLimits}.}
    \label{fig:paramSpace}
\end{figure}
It is instructive to compare this parametrically to the present-day abundance expected for the minimal model of a QCD axion with $\mathcal{O}(1)$ initial misalignment angle and zero-temperature mass $\maz$:
\begin{align}
    \f{\Omega_a^{\mathrm{(level \, cross)}}}{\Omega_{a}^{\mathrm{(minimal)}}} \sim \frac{\fd^2}{\fa^2} \p{\frac{\TQCD}{\sqrt{\mpl \md}}} \left(\frac{\mpl \maz}{ \TQCD^2} \right)^{\frac{2}{n+4}}\,,
\end{align}
where we have dropped numerical coefficients, and the terms inside parentheses are typically ${\mathcal O}({\rm few})$. Because the ratio $\fd/\fa$ can be large, it is clear that an adiabatic transfer of energy from an initial sterile field can provide a significant enhancement in the late-time abundance of the QCD axion.

In Fig.~\ref{fig:Evolution}, we show a representative example of these dynamics. We plot the axion energy densities (upper panel) and mass eigenvalues (lower panel) as functions of SM temperature, with temperature decreasing (time increasing) from left to right. At high temperatures, the heavy state comprises mostly the sterile axion, and the light state mostly the QCD axion. As the Standard Model plasma cools and QCD axion potential turns on, the eigenvalues approach one another and we observe an avoided crossing at $\Tcross$. In this example, $\Tcross$ occurs long after both axions have started oscillating so that the crossing takes place over many oscillations, therefore satisfying the condition of Eq.~\ref{eq:adiabaticityCondition}. As a consequence, the energy initially associated with the heavy state (and mostly with the sterile axion) remains with the heavy state. After the crossing, however, the heavy state mostly comprises the QCD axion, and its energy density increases as the universe cools due to the rapid increase in its mass. This carries on until the universe cools below $\TQCD$ and the relic QCD axion matter fraction is frozen in. In this example, parameters have been chosen so that the initial value of $\fd$ is precisely right to make the final energy density in the QCD axion equal to the present-day DM density.

We now explicitly check that the full range of QCD axion masses above the typical misalignment range $\maz\geq10\,\mu{\rm eV}$ are accessible through this mechanism. By setting $\Omega_a$ equal to the present-day observed DM abundance and taking a fiducial value of $\Theta_{0,S} = \pi/2$, we can solve Eq.~\ref{eq:relicAbundance} for the necessary $\fd$. Plugging this into Eq.~\ref{eq:adiabaticityCondition}, we find that the crossing will be adiabatic if:
\begin{align}\label{eq:adiabaticity}
    \f{10^6}{n}\p{\f{\md}{\rm eV}}^{\f{3}{4} + \f{4}{n}} \p{\f{\maz}{\rm eV}}^{-\p{\f{1}{2} + \f{4}{n}}}\gg 1\,.
\end{align}
In addition, we require $\fd \gg \fa$, since otherwise the mass eigenstates have a very different structure. Using the usual relation between $\fa$ and $\maz$ as well as the value for $\fd$ necessitated by Eq.~\ref{eq:relicAbundance}, we obtain the requirement:
\begin{align}
    6 \times 10^{5} \p{\f{\maz}{\rm eV}}^{1/2}\p{\f{\md}{\rm eV}}^{1/4}\gg 1\,.
\end{align}
We must also require that a crossing actually happens and that it happens while both axions have already started oscillating. This means we must require $\Tcross \ll \Tosc$ and $\md \ll \maz$. Computing $\Tcross$ and $\Tosc$ this first requirement reads:
\begin{align}\label{eq:Tx<Tosc}
    7\times 10^{-6}\p{\f{\md}{\rm eV}}^{-\p{\f{1}{2}+\f{2}{n}}} \p{\f{\maz}{\rm eV}}^{\f{2}{n}} < 1\,.
\end{align}
Finally, we must check that the value of $\fd$ required by Eq.~\ref{eq:relicAbundance} is not already ruled out by direct detection. This is a weaker requirement than the others, since it is possible that $\phid$ has no coupling to the SM, but we conservatively assume the presence of at least a sterile axion-photon coupling $\frac{g_{S \gamma \gamma}}{4} \phid F \tilde{F}$ with coupling strength $g_{S \gamma \gamma} \sim \alpha_{\rm QED}/(2 \pi \fd)$. This implies that the desired sterile axion would be ruled out unless the masses satisfy:
\begin{align}\label{eq:fD<astrophysics}
    5\times 10^{-5}\p{\f{\md}{\rm eV}}^{-\f{1}{4}} \p{\f{\maz}{\rm eV}}^{\f{1}{2}}\ll 1\,.
\end{align}
One can easily verify that these constraints may all be simultaneously satisfied over the entire high-mass QCD axion mass range of interest here: $10\,\mu{\rm eV}\leq\maz\leq1\,{\rm eV}$.

Fig.~\ref{fig:paramSpace} plots this available parameter space for QCD axions coupled to photons through $\f\gagg4 \phia \tilde F F$, with $F$ the electromagnetic field strength. The colored area is the QCD axion band, with the red region denoting the mass range accessible through misalignment with $\phia(0)/\fa\in[0.1\pi,0.9\pi]$ \cite{GrillidiCortona:2015jxo,Marsh:2015xka}, the blue region representing the expected range for post-inflationary production \cite{Buschmann:2021sdq}, and the green region indicating those parts of parameter space accessible with the adiabatic transfer described in this work. The dark gray regions represent excluded parameter space either by axion haloscopes \cite{ADMX:2018gho,ADMX:2018ogs,ADMX:2019uok,Crisosto:2019fcj,ADMX:2021nhd,ADMX:2021mio,Lee:2020cfj,Jeong:2020cwz,CAPP:2020utb,Lee:2022mnc,Kim:2022hmg,Yi:2022fmn,Adair:2022rtw,DePanfilis:1987dk,Hagmann:1990tj,HAYSTAC:2018rwy,HAYSTAC:2020kwv,HAYSTAC:2023cam,Alesini:2019ajt,Alesini:2020vny,Alesini:2022lnp,Quiskamp:2022pks} or by astrophysical probes \cite{Ayala:2014pea,Dolan:2022kul,Noordhuis:2022ljw}. In transparent light-gray, we plot the expected reach of some upcoming experiments  \cite{IAXO:2019mpb,Baryakhtar:2018doz,Schutte-Engel:2021bqm,BREAD:2021tpx,BRASS,Aja:2022csb,McAllister:2017lkb,Beurthey:2020yuq,Lawson:2019brd,DeMiguel:2023nmz,Stern:2016bbw,DMRadio:2022pkf,Berlin:2020vrk}. In particular, we point out that some experiments at higher masses (for example Refs.~\cite{Baryakhtar:2018doz,Schutte-Engel:2021bqm,BREAD:2021tpx}) are probing regions of parameter space which are generally not expected to be populated by the minimal formation mechanisms, but are naturally produced by the mechanism described here.

\section{UV Completion} \label{sec:UV}
The dynamics described in the previous section can arise in a wide variety of scenarios (see e.g. \cite{Kivel:2022emq}), and in this section we provide one concrete realization derived from the KSVZ mechanism \cite{Kim:1979if,Shifman:1979if}. Let $q_1$ and $q_{\rm mix}$ be new vector-like quarks charged under the Standard Model QCD gauge group, and let $q_2$ be a vector-like quark charged under some new dark confining gauge group. We now introduce two complex scalar fields $\Phi_1 = \rho_1 e^{\I \theta_1}$ and $\Phi_2 = \rho_2 e^{\I \theta_2}$, and suppose the Lagrangian is invariant under the following pair of $U_A(1)$ symmetries
\begin{gather}
    \Phi_1\to\Phi_1 e^{-\I\alpha}\,,\hspace{1cm}\Phi_2\to \Phi_2 e^{-\I\beta}\,,\\\notag
    q_1\to e^{\I\alpha\gamma^5/2}q_1\,,\hspace{0.4cm}q_{\rm mix}\to e^{\I\beta\gamma^5/2}q_{\rm mix}\,,\hspace{0.4cm}q_2\to e^{\I\beta\gamma^5/2}q_2\,,
\end{gather}
with $\alpha,\beta\in \R$.
As a consequence, the structure of the potential is limited to the form
\begin{align}\notag
    V &= \lambda_1 \Phi_1 \bar q_1 q_1 + \lambda_{\rm mix} \Phi_2 \bar q_{\rm mix} q_{\rm mix} + \\
    & + \lambda_2 \Phi_2 \bar q_2 q_2 + {\rm h.c.}+ V(|\Phi_1|^2,|\Phi_2|^2)\,.
\end{align}
Upon rotating away the complex phases of $\Phi_1$ and $\Phi_2$ in an axial rotation of the quarks, we are left with the following Lagrangian for the axion-(dark) gluon interactions
\begin{equation}
\begin{aligned}
    {\mathcal L}&\supset \f{1}{32\pi^2}(\theta_1 + \theta_2 + \arg\det M) G\tilde G \\&+ \f{1}{32\pi^2}(\theta_2 + \arg\det M_D) G_D\tilde G_D\,,
\end{aligned}
\end{equation}
where $M$ and $M_D$ are the (dark) quark mass matrices.
One may absorb both $\arg\det$'s into the choice of zero for $\theta_1$ and $\theta_2$. If the dark gauge group has an instanton condensate similar to the SM QCD gauge group, then the low-energy dynamics of this model will be the model of the previous section. Provided the dark gauge group confines long before QCD, then at the scales relevant for QCD it will lead to a temperature-independent potential for the sterile axion, giving precisely the low-energy Lagrangian shown in Eq.~\ref{eq:lagrangian}.

\section{Discussion} We have shown that adiabatic level crossing between a sterile axion and the QCD axion can lead to QCD axion dark matter at higher masses.
Our result provides motivation for several experiments that will probe this high-mass range in the coming years, and expands the known mechanisms by which the QCD axion may change its abundance. Other mechanisms that can produce heavy QCD axion DM often modify large or small scale galactic structure \cite{Co:2017mop,Harigaya:2019qnl,Co:2019jts,Kawasaki:2014sqa,Hiramatsu:2010yn,Hiramatsu:2012sc,Ringwald:2015dsf,Harigaya:2018ooc,Caputo:2019wsd}. In contrast, adiabatic level crossing relies only on linear dynamics, and therefore leaves the matter power spectrum largely unchanged (see e.g. Ref~\cite{Cyncynates:2021xzw}). Observation of a heavy QCD axion without matter power spectrum modifications is then suggestive of adiabatic transfer between a sterile axion and the QCD axion.

This is not the first time such adiabatic transfer has been noticed. In particular, previous work in Ref.~\cite{Ho:2018qur} (also see Ref.~\cite{Daido:2015bva,Daido:2015cba}) demonstrates a similar effect, wherein the QCD axion may transfer its energy to some sterile axion. The distinction between our mechanism and that of Ref.~\cite{Ho:2018qur} is particularly interesting: depending on whether the QCD axion couples to the sterile axion through the QCD instanton or the dark instanton potential, the energy flows either towards or away from the QCD axion respectively.\footnote{This distinction comes down to the misalignment initial conditions implied by the full non-linear potential.} This demonstrates not only the ease with which the QCD axion may change its abundance in a multi-axion theory, but also shows us that we may learn something about the structure of the multi-axion potential by measuring the abundance of the QCD axion.

\begin{acknowledgments}
The authors would like to thank Masha Baryakhtar,  Michael Fedderke, and Zachary Weiner for helpful comments on the manuscript.
D.C.\ is supported through the Department of Physics and College of Arts and Science at the University of Washington.
\end{acknowledgments}

\bibliographystyle{apsrev4-1}
\bibliography{bibliography}

\end{document}